\documentclass[twocolumn,aps,superscriptaddress,prb]{revtex4-2}
\usepackage{graphicx}
\usepackage{amssymb}
\usepackage{amsmath}
\usepackage{bm}
\usepackage{color}
\usepackage{braket}
\usepackage{siunitx}
\usepackage[top=30truemm,bottom=30truemm,left=25truemm,right=25truemm]{geometry}

\def\lsim{\, \lower -0.3ex \hbox{$<$} \kern -0.75em \lower 0.7ex \hbox{$\sim$} \,}
\def\gsim{\, \lower -0.3ex \hbox{$>$} \kern -0.75em \lower 0.7ex \hbox{$\sim$} \,}
\begin{document}

\title{
Effective continuum model of twisted bilayer GeSe and origin of emerging one-dimensional mode
}
\author{Manato Fujimoto}
\affiliation{Department of Physics, Osaka University,  Osaka 560-0043, Japan}
\author{Toshikaze Kariyado}
\affiliation{International Center for Materials Nanoarchitectonics (WPI-MANA), National Institute for Materials Science, Tsukuba 305-0044, Japan}
\date{\today}

\begin{abstract} 
The electric structure of twisted bilayer GeSe, which shows a rectangular moir\'{e} pattern, is analyzed using a $\bm{k}\cdot\bm{p}$ type effective continuum model. The effective model is constructed on the basis of the the local approximation method, where the local lattice structure of a twisted bilayer system is approximated by its untwisted bilayer with parallel displacement, and the required parameters are fixed with the help of the first-principles method. By inspecting the twist angle dependence of the physical properties, we reveal a relation between the effective potential under moir\'{e} pattern and the alignment of the Ge atoms, and also the resultant one-dimensional flat band, where the band is flattened stronger in a specific direction than the perpendicular direction. Due to the relatively large effective mass of the original monolayers, the flat band with its band width as small as a few meV appear in a relatively large angle. This gives us an opportunity to explore the dimensional crossover in the twisted bilayer platform. 
\end{abstract}

\maketitle
\section{Introduction}

Over the past decade, van der Waals (vdW) heterostructure has been offering a variety of rich phenomena to condensed matter and material science field \cite{geim2013van}. 
Incommensurate heterobilayers \cite{dos2007graphene,mele2010commensuration,trambly2010localization,PhysRevB.81.165105,morell2010flat,bistritzer2011moire,moon2012energy,de2012numerical,woods2014commensurate,cao2018unconventional,cao2018correlated,PhysRevX.8.031087,sunku2018photonic,yankowitz2019tuning,lu2019superconductors,jin2019observation,sharpe2019emergent,zondiner2020cascade,stepanov2020untying,uri2020mapping,shimazaki2020strongly,serlin2020intrinsic,zondiner2020cascade,wong2020cascade,nuckolls2020strongly,choi2021correlation} with small mismatch between layers lead to an interesting concept of moir\'{e} superlattices, where long-range pattern originating from the mismatch emerges to form a large unit cell. Amongst possible sources for the mismatch, a mismatch in relative angle between two layers is particularly interesting since it may allow us to change the material property as a function of the relative angle. Generically, a long-range potential induced by moir\'{e} pattern can affects the electrons in the heterostructures and can have a significant impact on its properties. 
For example, a flat band is formed in the electronic band dispersion in angle mismatched bilayer graphene, which is now called twisted bilayer graphene, at the magic angle, resulting in correlated insulating state, superconductivity \cite{cao2018unconventional,cao2018correlated,yankowitz2019tuning,lu2019superconductors} and fragile topology \cite{PhysRevX.8.031088,PhysRevLett.123.036401,PhysRevB.99.195455,PhysRevX.9.021013,song2021twisted,peri2021fragile}. Because the moir\'{e} pattern often gives an effective potential trapping electrons, twisted bilayer systems offer a promising stage to explore electron-electron interaction effects in a tunable manner. 

For interacting electrons, it is well known that the dimensionality of the system plays a crucial role.
In one-dimensional systems, electrons have no path to avoid each other, and thus, interact more strongly than in higher dimensional systems \cite{friesen1980dielectric,sarma1985screening,goni1991one,schulz1993wigner,tanatar1998wigner,altmann2001electronic,nagao2006one,hong2016one}.
One-dimensional systems provide an ideal platform for studying quantum many-body effects because their universal properties are described by the Tomonaga-Luttinger liquid theory applied to 1D models \cite{tomonaga1950remarks,luttinger1963exactly,haldane1981effective,voit1995one,giamarchi2003quantum,chang2003chiral}, showing novel phenomena such as spin-charge separation. Then, if we could realize a 1D system in some highly tunable system like a vdW heterostructure, it would give an excellent opportunity to study those intriguing phenomena.

Recently, a possible mechanism to realize (quasi) one-dimensional band dispersion in twisted bilayers is proposed in theoretical analysis of twisted bilayers with generic symmetry \cite{kariyado2019flat}. There, the one-dimensionality comes from anisotropic band flattening tuned by twist angles. The anisotropic band flattening is also predicted in twisted bilayer of GeSe, which is a typical group IV-monochalcogenide, through a large scale ab-initio simulation \cite{kennes2020one}. 
Around twisted angle $6^\circ$, it is found that the calculated band is nearly flat in one direction, while it is still dispersive in the perpendicular direction. The anisotropy is only moderate for large twist angles like more than 10$^\circ$, but it gets more and more prominent for smaller twist angles. 
That is, twisted bilayer GeSe provides a tunable playground for low-dimensional physics because of the crossover between two-dimensional and one-dimensional physics via varying the twisted angle. Notably, it also keeps the advantages of the two-dimensional materials, such as relatively easy carrier doping by gating.

Theoretically, there are several of possible strategies in analyzing properties of a twisted bilayer system. One possible way is to apply the first-principles method, typically the density functional theory (DFT), directly on a system with large unit cell. For convenience, we call this method \textit{direct method}. The direct method enables faithful simulations on a given target material within the accuracy and reliability of the first principles method, but its computational cost is usually high for a system with large unit cell. The prediction of the anisotropic band flattening in GeSe \cite{kennes2020one} has been done with the direct method with sophisticated large scale calculations. The other way, which complements the direct method, is to construct $\bm{k}\cdot\bm{p}$ type effective continuum models to describe low energy physics in twisted bilayers \cite{PhysRevB.86.155449}. When the angle mismatch is very small, a twisted bilayer is \textit{locally} well approximated by an untwisted bilayer \cite{PhysRevB.87.205404,PhysRevX.8.031087}. \textit{Globally}, the effect of the twist appears as a change of parallel displacement between two layers in the length scale of the moir\'{e} pattern. Then, scanning electronic structures of untwisted bilayers over all possible displacements allows us to obtain an effective model for the twisted bilayer. For convenience, we call this method to obtain an effective model \textit{local approximation}. The local approximation is usually good for having an intuitive understanding of the low energy physics, and once the effective model is obtained, it requires much less computational cost than the direct method. In this paper, we apply the local approximation to twisted bilayer GeSe to complement the direct method, and give an intuitive understanding of the anisotropic band flattening. 

The paper is organized as follows. First, we give a brief sketch of our local approximation scheme and a small remark on our computational method in Sec.~\ref{sec:method}. Then, we introduce the crystalline structure of twisted bilayer GeSe in Sec.~\ref{sec:crystal}, and we present an effective continuum model for the lowest conduction band of electric structure of twisted bilayer GeSe in Sec.~\ref{sec:continuum}.
Section~\ref{sec:band} is devoted for calculating the band structure of twisted bilayer GeSe. There, we demonstrate the modulation of one-dimensionality of the system in decreasing twist angle. The paper is concluded in Sec.~\ref{sec:summary} with a small discussion.

\section{Method}\label{sec:method}
Before going into details, we give a brief sketch of our local approximation scheme. As a first step, we derive crystalline structure parameters for monolayer GeSe. In the next step, we make the (untwisted) bilayer GeSe, and check how the interlayer distance depends on the displacement vector $\bm{\delta}$, whose definition will be clarified soon later. Note that if we apply the full structural optimization, the bilayer system automatically choose a $\bm{\delta}$ realizing the maximum binding energy (or it can be trapped at metastable structures). This does not fit our purpose, i.e., obtaining information over all possible $\bm{\delta}$. Therefore, we instead apply the rigid layer approximation where the crystalline structure for each layer is fixed and the optimized layer distance $d_z$ for each $\bm{\delta}$ is obtained by checking the layer distance dependence of the total energy with fixed $\bm{\delta}$. Then, using the obtained $d_z(\bm{\delta})$, we perform electronic structure analysis for all possible $\bm{\delta}$ to extract information required to construct the effective continuum model. 

During the course of analysis, we utilize Quantum Espresso package \cite{Giannozzi_2009,Giannozzi_2017} when the first-principles DFT calculation is required. For any calculations related to the crystalline structure analysis, we employ rev-vdW-DF2 type functional \cite{PhysRevB.89.121103} to take the van der Waals interaction into account, while for the other calculations, we employ PBE-GGA functional \cite{PhysRevLett.77.3865} for simplicity. The required pseudopotentials are taken from pslibrary \cite{pslibrary,DALCORSO2014337}. In order to simulate two-dimensional crystals, we use a unit cell that is sufficiently large in the direction perpendicular to the layer. 

\section{Crystlline Structure Analysis}\label{sec:crystal}
\subsection{Monolayer GeSe}

  \begin{figure*}[t]
  \begin{center}
    \leavevmode\includegraphics[width=1. \hsize]{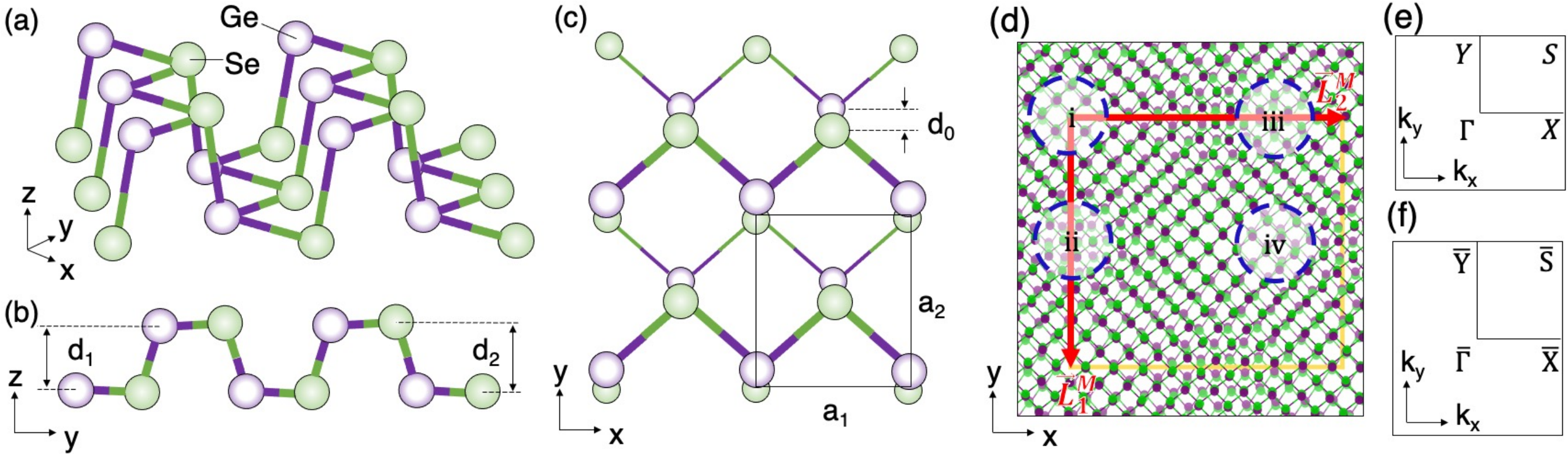}
 \caption{(a-c) The atomic structure of monolayer GeSe. (b) The projection to the $z$-$y$ plain. (c) The projection to the $x$-$y$ plain. The black solid box represents the unit cell. (d) Atomic structure and the moir\'{e} unit cell of twisted bilayer GeSe with $\theta=8^\circ$. The microscopic atomic structures in the region enclosed by the dashed circles, labeled by (i)...(iv), are illustrated in Figure \ref{fig_GeSe_bilayer}. (e,f) Brillouin zone of (e) monolayer and (f) twisted bilayer GeSe.}
    \label{fig_GeSe_structure}
  \end{center}
  \end{figure*} 

\begin{table}
\caption{The lattice parameters of monolayer GeSe and the atomic position of Ge and Se atoms in the unit cell obtained from the DFT calculation. The $xyz$ coordinates at the left are for monolayer GeSe, and those at the right are for the second layer to form bilayer GeSe.}\label{table_lattice}
\begin{tabular}[t]{|l|lll|lll|}
\hline
   & $x$      & $y$            & \multicolumn{1}{c|}{$z$} & $x$ & $y$ & $z$ \\ \hline
Ge & 0      & $d_0$         & $d_1$ & $\delta_1$ & $-d_0+\delta_2$ & $d_1+d_z$                 \\
Ge & $a_1/2$ & $a_2/2$ $+d_0$ & 0  & $a_1/2+\delta_1$ & $a_2/2-d_0+\delta_2$ & $d_z$                    \\
Se & 0      & 0            & 0    & $\delta_1$ & $\delta_2$ & $d_z$                  \\
Se & $a_1/2$ & $a_2/2$       & $d_1$    & $a_1/2+\delta_1$ & $a_2/2+\delta_2$ & $d_1+d_z$               \\ \hline
\end{tabular}
\vspace{5mm}
\begin{tabular}[t]{|l|l|l|l|l|}
\hline
$a_1$[nm] & $a_2$[nm] & \multicolumn{1}{c|}{$d_0$[nm]} & $d_1$[nm] & $d_2$[nm] \\ \hline
0.393  & 0.425  & 0.040                        & 0.250  & 0.254  \\ \hline
\end{tabular}
\end{table}
GeSe is a typical group-IV monochalcogenide material \cite{RevModPhys.93.011001} that crystallizes in a layered structure, where layers are bound mostly by week van der Waals interaction. Then, picking up a single layer, Ge and Se atoms are forming a rectangular lattice belonging to the space group $C_{2v}$.
Figure \ref{fig_GeSe_structure} shows the atomic structure of monolayer GeSe.
The purple atoms represent Ge and the green atoms represent Se.
The layer is placed on the $x$-$y$ plane, i.e., perpendicular to the $z$-axis and the origin is fixed at the position of one of the Se atoms. (See Table~\ref{table_lattice}.)

The layer is buckled, and the side view [Fig.~\ref{fig_GeSe_structure}(b)] shows an armchair type structure along the $y$-axis. On the other hand, the top view [Fig.~\ref{fig_GeSe_structure}(c)] shows a zigzag chain of Ge and Se in the $x$-direction. 
We obtain the lattice structure from DFT calculation and summarize the results in Table~\ref{table_lattice}. The results show a good agreement with the previous theoretical \cite{gomes2015phosphorene,C7CP07993A} and experimental \cite{doi:10.1021/ja107520b} works.

\subsection{Twisted bilayer GeSe}
Let us move on to the description of twisted bilayer GeSe. For this, we define $\bm{a}_1 = a_1(1,0,0)$ and $\bm{a}_2 = a_2(0,1,0)$ as the lattice vectors of the monolayer and $\bm{g}_i = 2\pi \bm{a}_i/(a_1a_2)$ as the associated reciprocal lattice vectors. We construct a twisted bilayer with relative rotation angle $\theta$ as a following manner. 
We first prepare two identical copies of layers, layer 1 and layer 2, at the same position, and apply the vertical shift only on layer 2, resulting in AA-stacking (untwisted) bilayer. Then, we rotate layers 1 and 2 with respect to the axis passing through the Se site at the origin by $-\theta/2$ and $\theta/2$, respectively. 
The lattice vectors of layer $l$ after the rotation are given by $\bm{a}^{(l)}_i = R(\mp \theta/2)\bm{a}_i$ with $\mp$ for $l = 1,2$, respectively, where $R(\theta)$ represents the rotation by $\theta$.
The corresponding reciprocal lattice vectors are $\bm{g}^{(l)}_i = R(\mp \theta/2)\bm{g}_i$ with $\mp$ for $l = 1,2$ as well.
The twist by $\theta$ and $-\theta$ are mirror images sharing equivalent band structures. 
The systems with twisted angles $\theta$ and $180^\circ-\theta$ form different moir\'{e} patterns.
The previous work \cite{kennes2020one} refers to them as configurations A and B, respectively, and reveal that the electric structure of configuration B exhibits the one-dimensional flat bands by DFT calculation.
Here, we will focus on configuration B in this work to study the one-dimensional flat bands and we refer to the twisted angle of $180^\circ-\theta$ as $\theta$ for simplicity.

When the rotation angle is small, the mismatch between the lattice vectors of the two layers gives rise to a long-range moir\'{e} pattern as shown in Fig~\ref{fig_GeSe_structure}(d).
Also in the small rotation angle limit, the local lattice structure near a certain point $\bm{r}$ approximates an untwisted bilayer GeSe with in plane relative displacement $\bm{\delta}$, which depends on the position as
\begin{align}
\bm{\delta}(\bm{r})&=[R(\theta / 2)-R(-\theta / 2)]\bm{r}\nonumber\\
&=2\sin{\frac{\theta}{2}}\hat{z}\times\bm{r}
\label{eq_displacement}
\end{align}
where $\bm{r}$ is measured from the rotation center. 
The period of the moir\'{e} pattern $\bm{L}_{i}^{M}$ can be obtained by the condition that $\bm{\delta}(\bm{L}_{i}^{M})$ coincides with a primitive lattice vector of the original AA-stacked bilayer. 
As Figure \ref{fig_GeSe_structure}(d) shows, we may choose the period as
\begin{equation}
\bm{L}_{i}^{M}=\frac{\bm{a}_i \times \hat{z}}{2\sin{\frac{\theta}{2}}}  \quad (i=1,2),
\label{eq_period}
\end{equation}
where we have used $\hat{z}\times(\bm{a}_i\times\hat{z})=\bm{a}_i$.
The corresponding moir\'{e} reciprocal lattice vectors satisfying $\bm{G}_{i}^{\mathrm{M}} \cdot \bm{L}_{j}^{\bm{M}}=2 \pi \delta_{i j}$ are written as
\begin{equation}
\bm{G}_{i}^{M}=2\sin{\frac{\theta}{2}}\bm{g}_i\times\hat{z}  \quad (i=1,2).
\end{equation}

\begin{figure}
  \begin{center}
    \leavevmode\includegraphics[width=1. \hsize]{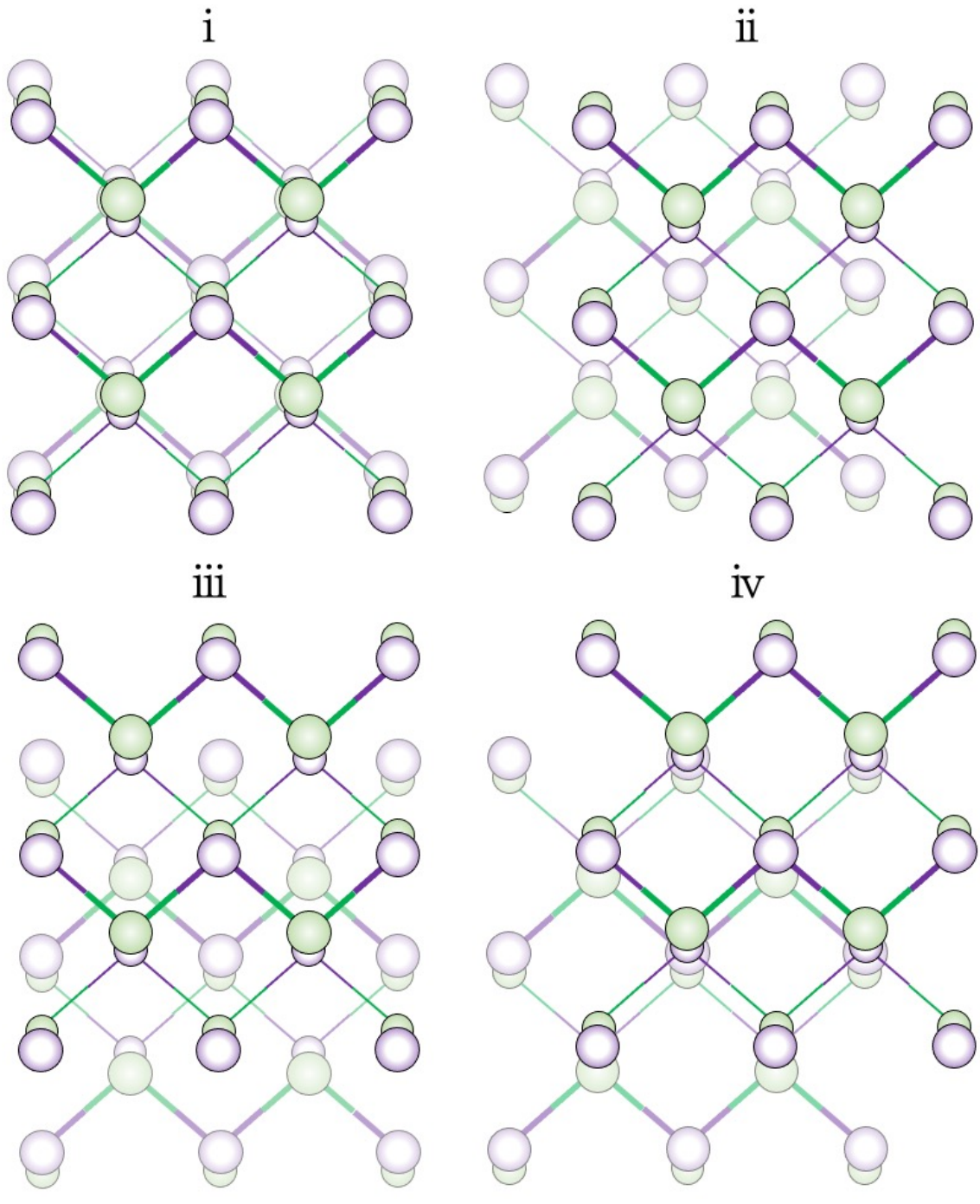}
 \caption{The atomic structure of untwisted bilayer GeSe with interlayer displacement $(r_1,r_2)=$ (i) $(0.02,0)$, (ii) $(1/2,0)$, (iii) $(0,1-2d_0/a_2)$ and (iv) $(1/2,1-2d_0/a_2)$. They are corresponding to the atomic structures at the labeled regions in Figure \ref{fig_GeSe_structure}(d).}
    \label{fig_GeSe_bilayer}
  \end{center}
  \end{figure}

Using Eqs.~\ref{eq_displacement} and \ref{eq_period}, $\bm{\delta}(\bm{r})$ at $\bm{r}=r_1\bm{L}^M_1+r_2\bm{L}^M_2$ can be written as
\begin{equation}
\boldsymbol{\delta}(\bm{r})=r_1 \bm{a}_1+r_2 \bm{a}_2.
\label{interlayer_slide_moire}
\end{equation}
Namely, near $\bm{r}=r_1\bm{L}^M_1+r_2\bm{L}^M_2$, the local structure is well approximated by the untwisted bilayer with the displacement of Eq.~\eqref{interlayer_slide_moire}. 
For example, the local crystalline structures in the four regions centered at $\bm{r}=r_1\bm{L}^M_1+r_2\bm{L}^M_2$ with $(r_1,r_2)=(0.02,0)$, $(1/2,0)$, $(0,1-2d_0/a_2)$, and $(1/2,1-2d_0/a_2)$ and labeled respectively by (i), (ii), (iii), and (iv) in Fig.~\ref{fig_GeSe_structure}(d) can be approximated by the crystalline structures of untwisted bilayer with displacement Eq.~\eqref{interlayer_slide_moire} with corresponding $(r_1,r_2)$ and labels shown in Fig.~\ref{fig_GeSe_bilayer}. Note that Eq.~\eqref{interlayer_slide_moire} also indicates that scanning over all possible $\bm{\delta}$ corresponds to scanning over the moir\'{e} unit cell.

\section{Effective continuum model}\label{sec:continuum}

\subsection{Formalisms}
Here we construct an effective continuum model for conduction band of electric structure of twisted bilayer GeSe where the emergence of one-dimensional modes is reported \cite{kennes2020one}.
The Hamiltonian of twisted bilayer system is generically written as \cite{dos2007graphene,mele2010commensuration}
\begin{equation}
H_{\mathrm{eff}}=\left(\begin{array}{cc}H_{1}(-i \nabla)+U(\bm{r}) & V(\boldsymbol{r}) \\ V^{\dagger}(\boldsymbol{r}) & H_{2}(-i \nabla)+U(\bm{r})\end{array}\right),
\label{eq_Ham_twisted_general}
\end{equation}
where $H_{l}$ are for the monolayer Hamiltonian in layer $l$ ($l=1,2$) and the off-diagonal components are for the interlayer tunneling. $U(\bm{r})$ can be induced by several of effects. For instance, it can be from the electrostatic potential from the partner layer. Or, the multiband effects can be reflected in $U(\bm{r})$. Namely, in constructing an effective model, we pick up a few of bands out of many bands in the original system, and the other bands can renormalize the potential. 

As we have discussed, a slightly twisted system is well approximated locally by untwisted bilayers, so the interlayer coupling in the twisted system can be derived from that of the untwisted system.
In the untwisted bilayer, no long-range superlattice is formed and the unit cell is spanned by the lattice vector for the original monolayer. (Note that 180$^\circ$ twist does not generate superlattices as well.)
Since the both layer has the same periodicity as the monolayer, the momentum in each layer is still a good quantum number, and the Hamiltonian for each momentum $\bm{k}$ in the original Brillouin zone is \cite{PhysRevB.95.245401}
\begin{equation}
H_{\bm{k}}=\left(\begin{array}{cc}H_0(\bm{k})+U_{\bm{k}}(\bm{\delta}) & V_{\bm{k}}(\boldsymbol{\delta}) \\ V_{\bm{k}}^{\dagger}(\bm{\delta}) & H_0(\bm{k})+U_{\bm{k}}(\bm{\delta})\end{array}\right).
\label{eq_Ham_untwisted}
\end{equation}
Just as in the case of Eq.~\eqref{eq_Ham_twisted_general}, $H_0(\bm{k})$ is for the Hamiltonian in each layer, and $V_{\bm{k}}(\bm{\delta})$ is for the interlayer coupling. $U_{\bm{k}}(\bm{\delta})$ can be again from the electrostatic potential from the partner layer or the multiband effect.
Generically, the interlayer coupling lifts the degeneracy caused by layer doubling. 
In the case that we can regard the electronic structure in each layer as a single band model, the eigenvalues $E_\pm(\bm{k})$ of Eq.~\eqref{eq_Ham_untwisted} becomes
\begin{equation}
    E^\pm_{\bm{k}}(\bm{\delta}) = H_{0}(\bm{k})+U_{\bm{k}}(\bm{\delta})\pm |V_{\bm{k}}(\bm{\delta})|. \label{eq:HUV}
\end{equation}
This means that if we know the energy gap $2\Delta_{\bm{k}}(\bm{\delta})$ caused by the layer degeneracy lifting, we can fix $V_{\bm{k}}(\bm{\delta})$ up to the phase as 
\begin{equation}
    |V_{\bm{k}}(\bm{\delta})|=\Delta_{\bm{k}}(\bm{\delta}).\label{eq:normV_delta}
\end{equation}
Microscopically, $V_{\bm{k}}(\bm{\delta})$ can also be derived by \cite{PhysRevB.86.155449,PhysRevB.87.205404,PhysRevX.8.031087}
\begin{equation}
V_{\bm{k}}(\bm{\delta})=\int d\bm{r}d\bm{r}'\psi_{\bm{k}}^{*}(\bm{r}+\bm{\delta})t_{12}(\bm{r}-\bm{r}')\psi_{\bm{k}}(\bm{r}'),
\label{eq_interlayer_hopping}
\end{equation}
where the area of the integral is the unit cell of monolayer, $\psi_{\bm{k}}(\bm{r})$ is Bloch wavefunction and $t_{12}$ is the microscopic interlayer hopping.

For constructing an effective model for the twisted bilayers, we want to relate $V_{\bm{k}}(\bm{\delta})$ with $V(\bm{r})$ and $U_{\bm{k}}(\bm{\delta})$ with $U(\bm{r})$. For this, we further assume that the electronic state in each layer ($l=1,2$) has quadratic dispersion and it is approximated as
\begin{equation}
H_{1/2}=E_{0}+\frac{1}{2} \hat{\bm{p}}_{1/2}M^{-1}(\mp\frac{\theta}{2}) \hat{\bm{p}}_{1/2}^\top,
\label{eq_Ham_single}
\end{equation}
where $\hat{\bm{p}}=-i \bm{\nabla}-R(\mp\theta/2)\bm{k}_{0}$.
The Hamiltonian is characterized by an inverse mass matrix $M^{-1}(\theta) $ around the energy $E_0$ and the momentum $\bm{k}_0$, where
\begin{equation}
  M^{-1}(\mp \frac{\theta}{2}) = R(\mp\frac{\theta}{2})\left(
    \begin{array}{cc}
      m_{xx}^{-1} & 0  \\
      0 & m_{yy}^{-1}
    \end{array}
  \right)R(\pm\frac{\theta}{2}).
\label{eq_inverse_mass}
\end{equation}
 In the small angle limit, the moir\'{e} Brillouin zone gets folded, and we usually see a few bands in a certain small energy range. 
 If we focus on a small energy window around $E_0$, Eq.~\eqref{eq_Ham_single} indicates that the relevant states are mostly contributed from the states with $\bm{k} \sim \bm{k}_0$ in each layer.
 In such a case, we expect that $V(\bm{r}) \sim V_{\bm{k}_0}(\bm{\delta}(\bm{r}))$ and $U(\bm{r}) \sim U_{\bm{k}_0}(\bm{\delta}(\bm{r}))$ form a good approximation \cite{PhysRevX.8.031087}.

Later it turns out that we can set $\bm{k}_0$ at the $\Gamma$-point for the twisted bilayer GeSe. 
In such a case, we can determine the interlayer coupling in Eq.~\eqref{eq_interlayer_hopping} as a real function because we can safely choose the gauge such that $\psi_{\Gamma}^{*}(\bm{r})=\psi_{\Gamma}(\bm{r})$ from the time reversal symmetry. (We have naively assumed that $t_{12}(\bm{r})$ is a real funciton.) Then, Eq.~\eqref{eq:normV_delta} fixes $V_\Gamma(\bm{\delta})$ up to its \textit{sign}, more than its \textit{phase}. Usually, $V_\Gamma(\bm{\delta})$ is a smooth function of $\bm{\delta}$, and if $\Delta_\Gamma(\bm{\delta})$ never hits zero, we can write
\begin{equation}
    V(\bm{r}) \sim V_\Gamma(\bm{\delta}(\bm{r})) = \Delta_{\Gamma}(\bm{\delta}(\bm{r})).\label{eq:V_delta}
\end{equation}

\subsection{Deriving parameters}
  Summarizing the above results, what we need to do is to derive $m_{xx}$, $m_{yy}$, $U_\Gamma(\bm{\delta})$, and $\Delta_{\Gamma}(\bm{\delta})$, and we do this with the help of the first-principles method. 
For this purpose, we calculate the band structure of the untwisted bilayer GeSe in DFT calculation with various interlayer displacements $\bm{\delta}$. 

\begin{figure}
    \centering
    \includegraphics[width=\hsize]{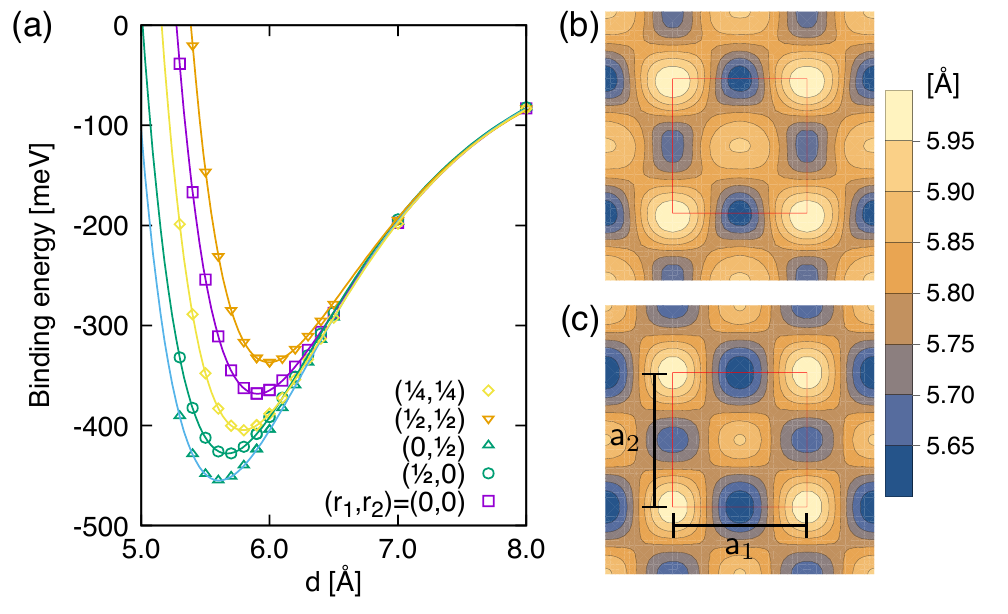}
    \caption{(a) Binding energy as a function of the interlayer distance for the selected $\bm{\delta}$. (b) Optimized interlayer distance obtained from DFT. (c) Fitting to the DFT results.}
    \label{fig:layer_distance}
\end{figure}

As a preparation, we first derive $\bm{\delta}$ dependence of the interlayer distance. Generically in a twisted bilayer system, its interlayer distance can depend on position, i.e., there is corrugation. Again, adapting the approximation that the local structures can be well described by the untwisted bilayers, this position dependence can be captured by the $\bm{\delta}$ dependence of the interlayer distance in the untwisted system. The crystalline coordinates with 180$^\circ$ rotation, displacement $\bm{\delta}$, and the interlayer distance $d_z$ are summarized in Table~\ref{table_lattice}. Using this coordinate set, we derive the optimized $d_z$ for each $\bm{\delta}$ by varying $d_z$ and maximizing the size of the binding energy within the DFT calculation. In the actual calculations, we use a unit cell whose c-axis length is 30 \AA, and define the binding energy as a difference between the total energy for given $d_z$ and $\bm{\delta}$ and the total energy of a reference state specified by $d_z=15$ {\AA} and $\bm{\delta}=0$ \cite{Hsing_2014}. Note that $d_z=15$ {\AA} is the largest possible layer distance for the unit cell with c-axis length being 30 \AA, and with this large distance, $\bm{\delta}$ dependence of the total energy is negligible. In order to extract $d_z$ that realizes the maximum size of the binding energy for each $\bm{\delta}$, we fit the calculated data by 
\begin{equation}
    E_{\text{bind}}(d_z)=\alpha \exp(-\beta(d_z-d_z^{(0)})) - \gamma(d_z^{(0)}/d_z)^{\zeta} \label{eq:binding_energy_fit}
\end{equation}
with $d_z^{(0)}=6$ \AA. Figure~\ref{fig:layer_distance}(a) shows the $d_z$ dependence of the binding energy for the selected $\bm{\delta}$s, and we can see that the fitting by Eq.~\eqref{eq:binding_energy_fit} is working well. The obtained optimal interlayer distance as a function of $\bm{\delta}$ is given in Fig.~\ref{fig:layer_distance}(b). To obtain the map in Fig.~\ref{fig:layer_distance}(b), the optimized $d_z$ is derived on the 18$\times$18 regular grid in the unit cell. For later convenience, the $\bm{\delta}$ dependence of the interlayer distance is approximated by a simple function
\begin{multline}
 d_z(\bm{\delta})=d_0+d_1\cos\tilde{r}_1+d_2\cos\tilde{r}_2\\
 +d_3(\cos(\tilde{r}_1+\tilde{r}_2)+\cos(\tilde{r}_1-\tilde{r}_2))
\end{multline}
with $\tilde{r}_i=2\pi r_i$. The choice of $\{d_0,d_1,d_2,d_3\}=\{5.795,-0.038,-0.007,0.078\}$ {\AA} results in the approximated map in Fig.~\ref{fig:layer_distance}(c), and we can see that the approximation gives a satisfactory match to the original data. Strictly speaking, the map in Fig.~\ref{fig:layer_distance}(b) does not have 180$^\circ$ rotation symmetry due to the symmetry of the GeSe layer, while the map in Fig.~\ref{fig:layer_distance}(c) does due to the simplicity of the approximation, but the deviation is small. 

\begin{figure}
  \begin{center}
    \leavevmode\includegraphics[width=1. \hsize]{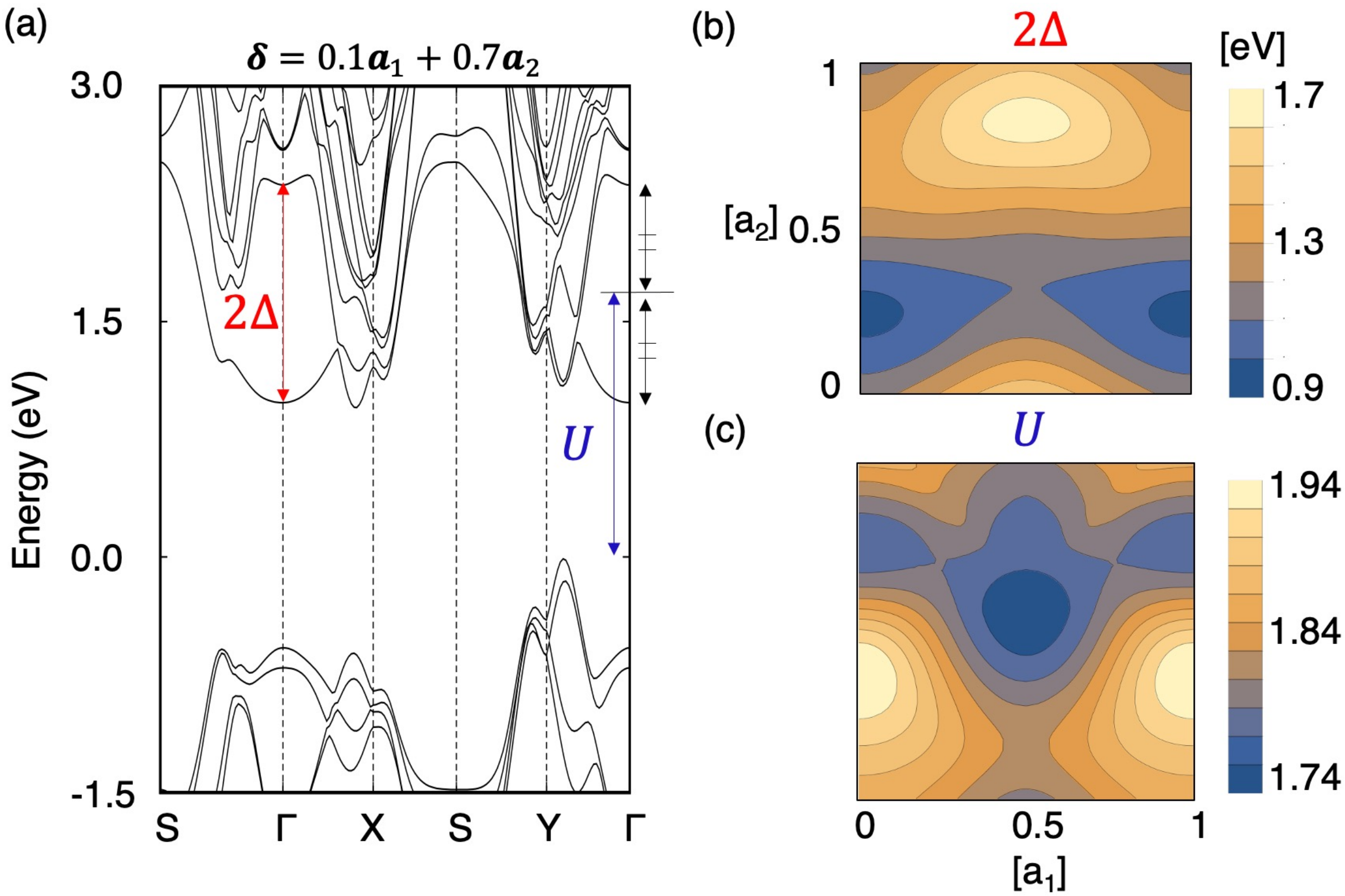}
 \caption{(a) The band structure of the untwisted bilayer GeSe with the interlayer displacement $\boldsymbol{\delta}=0.1\bm{a}_1+0.7\bm{a}_2$. $2\Delta$ is denoted as the energy difference between the first and second band from the bottom of the conduction band and its $\bm{\delta}$ dependence is plotted in (b). $U$ represents the average energy of the first and second lowest conduction band, and its $\bm{\delta}$ dependence is plotted in (c).}
    \label{fig_potential_vs_delta}
  \end{center}
  \end{figure}

Having the displacement dependence of the interlayer distance $d_z(\bm{\delta})$, we now inspect the $\bm{\delta}$ dependence of the electronic band structures for the untwisted bilayers. As we scan over $\bm{\delta}$, we build a system using $d_z(\bm{\delta})$ and perform DFT calculations for the band structures. As an example, Fig.~\ref{fig_potential_vs_delta}(a) shows the band structure at $\bm{\delta}=0.1\bm{a}_1+0.7\bm{a}_2$. From the scan over $\bm{\delta}$, we notice that the lowest energy in the conduction band is realized at the $\Gamma$-point, and thus, we choose to build an effective model around the $\Gamma$-point. Choosing the lowest energy conduction band as a target state means that we work on the slightly electron doped samples. Note that the minimum of the lowest conduction band for the monolayer is away from the $\Gamma$-point, but the interlayer tunneling effect is large at the $\Gamma$-point and then the energy of the lowest conduction band at the $\Gamma$-point becomes lowest among the conduction band at a certain $\bm{\delta}$. Targeting at the $\Gamma$-point, we derive $m_{xx}$ and $m_{yy}$ by inspecting the curvature of the lowest conduction band at the $\Gamma$-point obtained in the monolayer DFT band calculation. The obtained values are $m_{xx}^{-1}=0.34$, $m_{yy}^{-1}=0.5$ in the unit of the inverse mass of the free electron, which give $\frac{\hbar^2}{2m_{xx}}\bigl(\frac{\pi}{a_1}\bigr)^2\sim 0.83$ eV and  $\frac{\hbar^2}{2m_{yy}}\bigl(\frac{\pi}{a_2}\bigr)^2=1.04$ eV.

In order to derive $V_\Gamma(\bm{\delta})$ and $U_\Gamma(\bm{\delta})$, we investigate the energies of the first and the second lowest conduction band at the $\Gamma$-point for each $\bm{\delta}$, $E_{\Gamma 1}(\bm{\delta})$ and $E_{\Gamma 2}(\bm{\delta})$. By deliberately shifting the interlayer distance from the optimized value, specifically from the optimized distance to far away, we observe that the splitting between $E_{\Gamma 1}(\bm{\delta})$ and $E_{\Gamma 2}(\bm{\delta})$ gradually goes to zero, which indicates the splitting is basically caused by the interlayer tunneling. 
Then, from the difference and the average of $E_{\Gamma 1}(\bm{\delta})$ and $E_{\Gamma 2}(\bm{\delta})$, we can extract $U_\Gamma(\bm{\delta})$ and $\Delta_\Gamma(\bm{\delta})$ using Eqs.~\eqref{eq:HUV} and \eqref{eq:normV_delta}. (Strictly speaking, what we extract is $E_0+U_\Gamma(\bm{\delta})$, but a constant shift of energy is not particularly important.)
Figures~\ref{fig_potential_vs_delta}(b) and \ref{fig_potential_vs_delta}(c) show the obtained $\Delta_\Gamma(\bm{\delta})$ and $U_\Gamma(\bm{\delta})$. We first notice that $\Delta_\Gamma(\bm{\delta})$ never hits zero, suggesting that the sign of $V_\Gamma(\bm{\delta})$ is uniform and enabling us to set $V_\Gamma(\bm{\delta})=\Delta_\Gamma(\bm{\delta})$ as we have noted.
Importantly, the regions of high magnitude of interlayer coupling are distributed in a one-dimensional manner.

The amplitude of the interlayer coupling is related to the relative positions of Ge in layer 1 and the lower Ge because the conduction band minimum is dominated by Ge orbitals \cite{liu2021antibonding}.
For example, the interlayer coupling is maximal at $(r_1,r_2)=(1/2, 1-2d_0/a_2)$, where the corresponding atomic structure is shown in Fig.~\ref{fig_GeSe_bilayer}(iv), and we can see that one of the Ge atoms in layer 2 comes right on top of one of the Ge atoms in layer 1.
Starting from this structure, if the layer 2 is slid in $\pm y$ direction, we expect that the overlap of the Ge orbitals between layer 1 and 2 decreases rapidly. This explains the anisotropy or the one-dimensionality of the interlayer coupling.

From the all above, the continuum model in Eq.~\eqref{eq_Ham_twisted_general} can be written as
\begin{equation}
H =\left(\begin{array}{cc}H_{1}(-i \nabla) + U(\bm{\delta}(\bm{r})) & \Delta(\bm{\delta}(\bm{r})) \\ \Delta(\bm{\delta}(\bm{r})) & H_{2}(-i \nabla) + U(\bm{\delta}(\bm{r})) \end{array}\right)
\label{eq_Ham_effective_tbGeSe}
\end{equation}
and hence the eigenstates of the twisted bilayer GeSe can be obtained by diagonalizing the Hamiltonian matrix.
   \begin{figure*}[t]
  \begin{center}
    \leavevmode\includegraphics[width=1. \hsize]{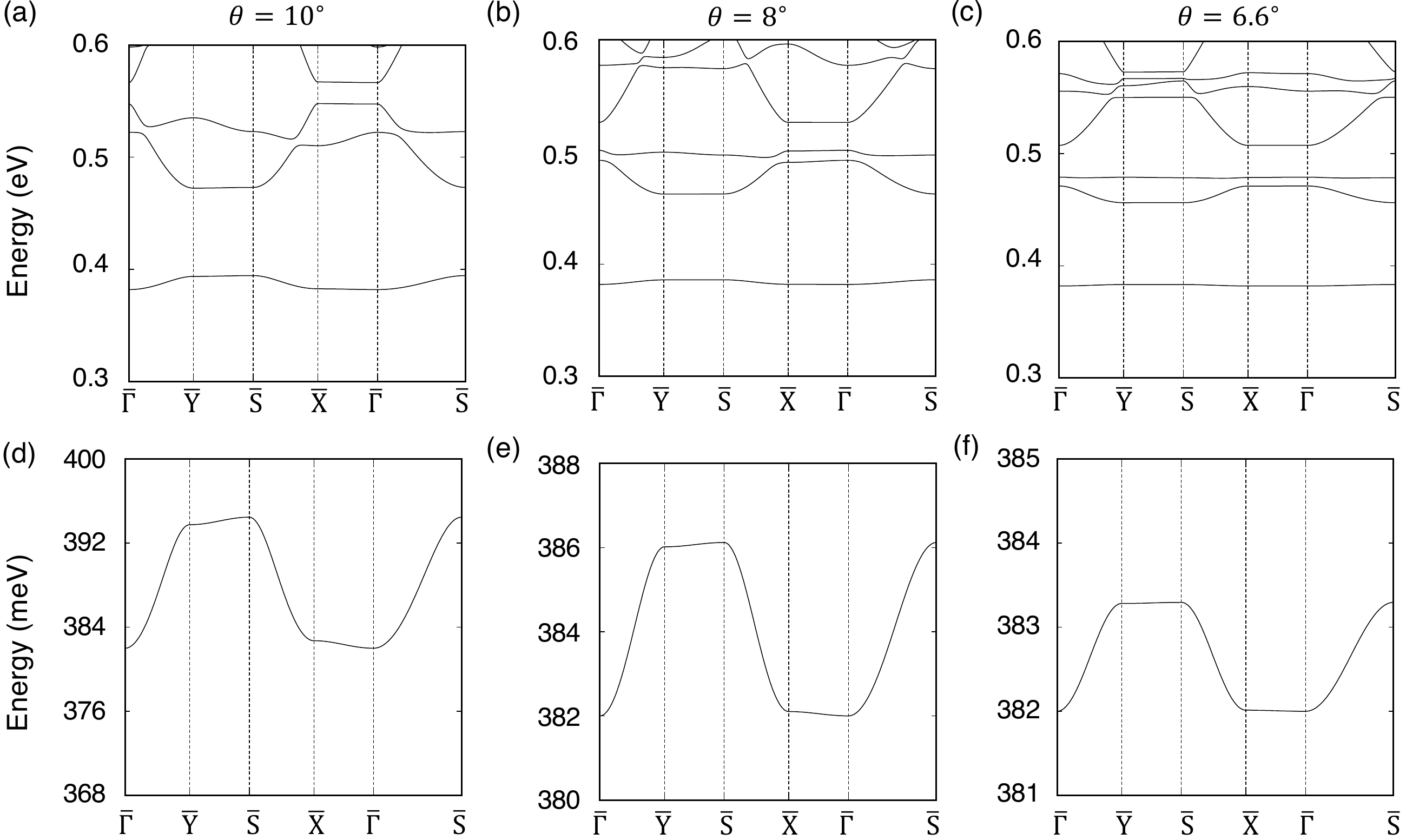}
 \caption{Band structure at twisted angle (a,d) $\theta=10^{\circ}$, (b,e) $\theta=8^{\circ}$, and (c,f) $\theta=6.6^{\circ}$. The panels (d), (e) and (f) are the close up view of the band structures in (a), (b) and (c), respectively.}
    \label{fig_band_theta}
  \end{center}
  \end{figure*}

\section{Electronic Structure}\label{sec:band}
  
   \begin{figure}
  \begin{center}
    \leavevmode\includegraphics[width=1. \hsize]{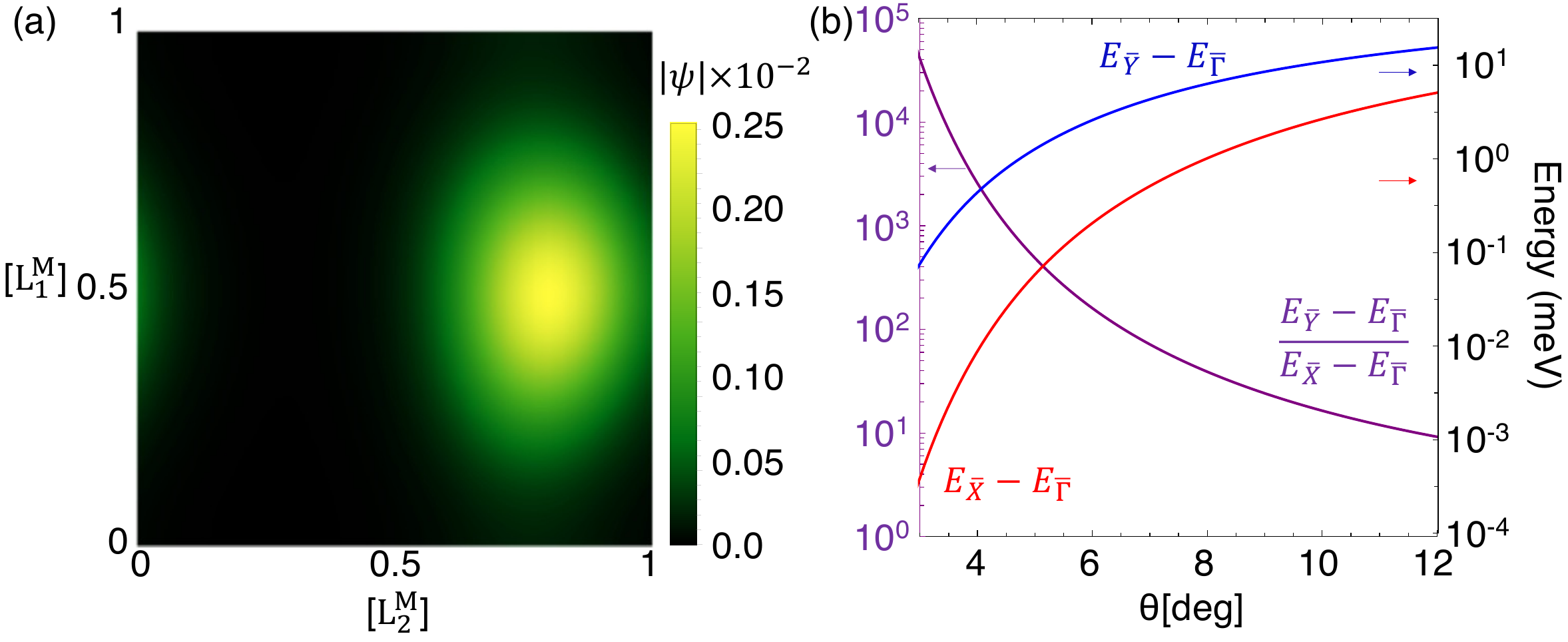}
 \caption{(a) The calculated wave function at $\bar{\Gamma}$ point for $\theta = 6.6^{\circ}$. (b) The red (blue) line shows the difference of energy between $\bar{X}$ ($\bar{Y}$) and $\bar{\Gamma}$ point, $E_{\bar{X}}-E_{\bar{\Gamma}}$ ($E_{\bar{X}}-E_{\bar{\Gamma}}$), as a function of $\theta$, while the purple line represents the ratio of $E_{\bar{X}}-E_{\bar{\Gamma}}$ to $E_{\bar{Y}}-E_{\bar{\Gamma}}$.}
    \label{fig_band_velocity_psi}
  \end{center}
  \end{figure} 
  
Now, we calculate the electronic band structures for the twisted bilayer GeSe by solving Eq.~\eqref{eq_Ham_effective_tbGeSe}. Numerically, it is solved by the plane wave expansion method. The angle dependence is encoded in the rotation matrix in $H_{1,2}$ and Eq.~\eqref{eq_displacement}. 

Figure~\ref{fig_band_theta} shows the calculated energy bands of the twisted bilayer GeSe with different rotation angles, $\theta = 10^{\circ}$, $8^{\circ}$ and $6.6^{\circ}$.
The horizontal axes are labeled by the high symmetric points of the Brillouin zone for the moir\'{e} superlattice [see Fig.~\ref{fig_GeSe_structure}(f)].
The band structures are similar for all twisted angles, but the overall energy scale rapidly shrinks as the rotation angle decreases.
There are the three isolated bands in the shown energy range for these angles. For instance, there opens an energy gap about $80$ meV between the lowest conduction band and the first excited band at $\theta=10^\circ$, and in decreasing $\theta$, we see that the energy gap gets smaller.
For the shown isolated bands, the band dispersions in $k_x$ direction is extremely smaller than that of in $k_y$ direction. 
Figure~\ref{fig_band_velocity_psi}(a) shows the calculated wavefunction at the $\bar{\Gamma}$-point for $\theta=6.6^\circ$. 
The wave function is localized where the interlayer coupling is maximum (note that the direction of $\bm{L}^{M}_i$ is 90$^\circ$ degree rotated from the direction of $\bm{a}_i$.) and spreads out in the $y$ direction.

Here, we focus on the band dispersion of the lowest conduction band for which the emergence of one-dimensional flat band is predicted thorough DFT calculation \cite{kennes2020one}.
Figure \ref{fig_band_velocity_psi}(b) shows the difference of the energy between $\bar{X}$ ($\bar{Y}$) and $\bar{\Gamma}$ point, i.e., $E_{\bar{X}}-E_{\bar{\Gamma}}$ ($E_{\bar{Y}}-E_{\bar{\Gamma}}$), by the red (blue) line and it measures the size of dispersion along $x$ ($y$) direction. The purple line in Fig.~\ref{fig_band_velocity_psi}(b) represents the ratio of $E_{\bar{X}}-E_{\bar{\Gamma}}$ to $E_{\bar{Y}}-E_{\bar{\Gamma}}$. While both of $E_{\bar{X}}-E_{\bar{\Gamma}}$ and $E_{\bar{Y}}-E_{\bar{\Gamma}}$ rapidly decays in decreasing twist angle, the ratio of the dispersion in $x$ to $y$ direction rapidly grows, indicating that we can modulate the strength of the one-dimensionality by controlling the twist angle.

\section{Discussions and Summary}\label{sec:summary}

In this paper, we analyzed the electric structure of twisted bilayer GeSe.
By calculating band structures of the effective model of the conduction band bottom derived from the local approximation method, we revealed that the one-dimensional flat band appears in accordance with the anisotropy of the interlayer potential originating from the alignment of Ge atoms. Note that we focus on the electronic states originally from the $\Gamma$-point, which means that the interlayer coupling is determined by the microscopic details because it is not constrained by the interference of the Bloch phase \cite{kariyado2019flat}. But in turn, we could fix $V(\bm{r})$ by $\Delta_\Gamma(\bm{\delta})$ without ambiguity in the phase. Our results indicate that the band width as small as meV order is reached in relatively large twist angle. This is mostly due to the relatively large effective mass in the original monolayer.

Now, let us compare the results in this paper (the local approximation) and the previous paper (the direct method) \cite{kennes2020one}. In comparing the results, we should be very careful that the used unit cells are different. Generically, a moir\'{e} unit cell is not always matching to the actual unit cell in the microscopic viewpoint. Namely, two regions connected by the moir\'{e} unit vector may possibly be different microscopically. In this paper, we simply use the moir\'{e} unit cell as a unit cell, since the microscopic details are washed out in building the continuum model. On the other hand, the strict periodicity in the microscopic sense is important in the direct method, and it appears that a unit cell 2$\times$2 larger than here was used. 

Having this difference in the unit cell in mind, the band structure calculated from the effective model in this paper looks like showing stronger flattening than that of the direct method \cite{kennes2020one}.  In this paper, we totally rely on the local approximation, and we speculate that it may probably overestimates the effective interlayer coupling.
The small change in the interlayer coupling can potentially cause a large effect to the flatness of the band because the band width is decaying exponentially in decreasing twisted angle. 

Another thing we should note is that the electron-electron interaction effects are neglected in analyzing the effective model. Of course, in deriving the required parameters, we have used the first-principles method, in which the electron-electron interaction is taken into account at the level of the modern density functional theory. However, once the effective model is obtained, we simply solve that $2\times 2$ Schr\"{o}dinger type equation using the plane wave basis. The construction of the elaborated theoretical framework, which may include the electron-electron interaction effects in a better way, is left for the future work. 

Another possible reason for the deviation from the direct method is related to the lattice relaxation. In our treatment, the corrugation effects are taken into account, while in-plane lattice relaxation effects are neglected. Namely, we adapted the rigid layer approximation in deriving the optimal layer distance, and dropped possible deformation within each layer from the consideration. On the other hand, in Ref.~\cite{kennes2020one}, they perform the full lattice relaxation. However, we speculate that the in-plane lattice relaxation is not the primary reason for the discrepancy. Namely, from the recent study of twisted bilayer graphene, it is suggested that the in-plain lattice relaxation is important in very small angle typically smaller than 2$^\circ$ \cite{Dai_2016,PhysRevB.96.075311}, which is smaller than the typical angle treated in this paper. However, we have to perform more works specific to GeSe to be more conclusive. 

Lastly, the microscopic details washed out in constructing the continuum model may also be important. It is a standard fashion to neglect the microscopic details in continuum models for moir\'{e} systems and we follow that fashion here, but since the energy scale of the focused band is very small, we may have to include subtle changes caused by the microscopic details. Overall, for the future developments, it is very important to have mutual feedback between the direct method, the local approximation, and experiments.

\section*{Acknowledgments}
This work was supported by JSPS KAKENHI Grant Number JP20K03844. M.F. was supported by a JSPS Fellowship for Young Scientists. Part of the computations in this work has been done using the facilities of the Supercomputer Center, the Institute for Solid State Physics, the University of Tokyo. 
\bibliography{Twisted_bilayer_GeSe.bib}
\end{document}